\documentclass[prd,twocolumn]{revtex4}%
\usepackage{amsfonts}
\usepackage{amsmath}
\usepackage{amssymb}
\usepackage{graphicx}%
\begin{document}
\title{Enhanced anti-Unruh effect by simulated light-matter interactions}
\author{Yongjie Pan}
\author{Baocheng Zhang}
\email{zhangbaocheng@cug.edu.cn}
\affiliation{School of Mathematics and Physics, China University of Geosciences, Wuhan
430074, China}

\begin{abstract}
We investigate the transition of a two-level atom as the Unruh-Dewitt detector
accelerated in the electromagnetic field in this paper. The enhancement of the
transition probability is found for different field states under the
conditions that the anti-Unruh effect appears. In particular, the enhancement
is the most prominent when the field state takes the squeezed state. We also
construct a new atomic trajectory to realize the effect of
acceleration-induced transparency. Based on this, we show that the anti-Unruh
effect still exists even at the specific atomic energy difference where
acceleration-induced transparency happens.

\end{abstract}
\maketitle

\section{Introduction}

The Unruh effect \cite{unruh1976notes,unruh1984happens} states that the
uniformly accelerated observers can feel a thermal bath of particles in the
Minkowski vacuum of a free quantum field. In the past few decades, it has been
extensively investigated in many different situations (see the review
\cite{crispino2008unruh} and references therein). An important application of
the Unruh effect is connected with the Unruh-DeWitt (UDW) detector
\cite{Hawking1979GeneralRA} which consists of a two-level atom accelerated in
the vacuum. When such an atom is accelerated, it feels a thermal bath
\cite{bell1983,vanzella2001decay,schutzhold2006signatures,oshita2016quantum,y2016non,cozzella2017proposal,lynch2021experimental,blasone2020neutrino}%
, which finally leads to a thermal balance between the accelerated atom and
the vacuum field satisfying the Kubo-Martin-Schwinger (KMS) condition by the
change of the probability distribution in two different energy levels of the
accelerated detector
\cite{feng2022quantum,takagi1986vacuum,kubo1986brownian,martin1959theory,fewster2016waiting}%
. However, the Unruh effect has not been observed directly in any experiments
up to now due to the pretty low Unruh temperature, e.g., it requires an
acceleration of $10^{20}m/s^{2}$ to realize a thermal bath at $1$ K.

A recent work \cite{vsoda2022acceleration} suggested a potentially strong
enough method to detect the Unruh effect. In their method, the UDW detector is
accelerated in the electromagnetic field instead of the vacuum. Thus, not only
the conventional Unruh effect (the counter-rotating terms) is activated but
also the rotating-wave terms in the usual light-matter interaction
\cite{cohen2004atoms} are also influenced by the acceleration. For the Fock
state, the Unruh effect is enhanced by a factor of $n+1$ ($n$ is the photon
number) compared with the case that happened in the vacuum. But the
enhancement can only be observed in some specific energy differences of the
accelerated atom, in which the disturbance of rotating-wave terms is much
suppressed and acceleration-induced transparency appears.

The recently found anti-Unruh effect
\cite{brenna2016anti,garay2016thermalization} states that a particle detector
in the uniform acceleration coupled to the vacuum can cool down with
increasing acceleration under certain conditions, which is opposite to the
celebrated Unruh effect. Under the situation of the anti-Unruh effect, some
novel and interesting phenomena were found and studied
\cite{li2018would,pan2020influence,pan2021thermal,zhou2021massive,barman2021radiative,chen2022entanglement,yan2022spacetime}%
. Although the anti-Unruh effect can lead to different behaviors for the
change of entanglement among many atoms from that led by the usual thermal
effect \cite{pan2021thermal}, no feasible methods in the current technical
conditions are found to detect it. An instant thought is whether the
anti-Unruh effect could be enhanced by using the accelerated detector in the
electromagnetic field along the line in Ref. \cite{vsoda2022acceleration}.
However, the occurrence of this enhancement requires generating the
acceleration-induced transparency for the accelerated atom in the
electromagnetic field, which is dependent on the atomic energy difference as
stated above. While the appearance of the anti-Unruh effect is also closely
related to the energy difference of the accelerated atoms
\cite{garay2016thermalization}, it is unclear whether the anti-Unruh effect
can appear when the acceleration-induced transparency happens for an
accelerated atom in the electromagnetic field and should be investigated
further. In this paper, we investigate this in detail. We consider the
different field states, including the Fock state, the thermal state, the
coherent state, and the squeezed state, and study whether the anti-Unruh
effect can appear in these field states and whether the anti-Unruh effect and
acceleration-induced transparency can appear under the same conditions.

This paper is organized as follows. In the second section, we describe the
interaction between the UDW detector and the field, and compare the results
for the accelerated detector in the vacuum with that in the electromagnetic
field. We study the stimulated anti-Unruh effect for the different field
states in the third section. This is followed in the fourth section by
choosing a specific accelerated trajectory to suppress the change of atomic
state induced by the light-matter interaction. This is the effect of
acceleration-induced transparency. We show that acceleration-induced
transparency and the anti-Unruh effect can appear under the same conditions.
Finally, we summarize and give the conclusion in the fifth section. In this
paper, we use units with $c=\hbar=k_{B}=1$.

\section{The Model}

We start with the model of the UDW detector which is considered as a pointlike
two-level atom in this paper. It has two different energy levels, described by
the ground $|g\rangle$ and excited $|e\rangle$ states, respectively. The two
energy levels are separated in the atomic rest frame by an energy gap $\Omega
$. When the atom is accelerated, the interaction Hamiltonian is expressed as
\cite{dewitt1979quantum,louko2008transition}
\begin{equation}
H_{I}=\lambda\chi(\tau)\mu(\tau)\phi\lbrack x(\tau),t(\tau)], \label{H}%
\end{equation}
where $\lambda$ is the coupling constant between the accelerated atom and the
field, $\mu(\tau)=e^{i\Omega\tau}\sigma^{+}+e^{-i\Omega\tau}\sigma^{-}$ is the
atomic monopole moment ($\sigma^{\pm}$ being SU(2) ladder operators),
$\phi\lbrack x(\tau),t(\tau)]$ is the field operator in which $x(\tau
),t(\tau)$ represents the trajectory of the atom, and $\chi(\tau)$ is the
switching function. In this paper, we choose $\chi(\tau)$ to be Gaussian
function
\begin{equation}
\chi(\tau)=e^{-\frac{\tau^{2}}{2\varsigma^{2}}},
\end{equation}
where the parameter $\varsigma$ establishes the timescale of the interaction
between the field and the detector. According to the earlier analyses
\cite{garay2016thermalization,pan2021thermal}, the interaction time will not
affect the existence of the anti-Unruh effect. Thus, we can extend the
interaction time to infinity, and the anti-Unruh phenomenon will still exist.

The time evolution operator under the Hamiltonian (\ref{H}) is obtained by the
following perturbative expansion,
\begin{align}
U  &  =1-i\int d\tau H_{I}(\tau)+\mathcal{O}(\lambda^{2})\nonumber\\
&  =1-i\lambda\sum_{k}[\eta_{(+,k)}a_{k}^{\dag}\sigma^{+}+\eta_{(-,k)}%
a_{k}\sigma^{+}+H.c.] \label{eu}%
\end{align}
where $a_{k}^{\dag}\sigma^{+}$ and $a_{k}\sigma^{-}$ are the counter-rotating
terms, while $a_{k}\sigma^{+}$ and $a_{k}^{\dag}\sigma^{-}$ are the
rotating-wave terms. $\sum_{k}$ represents the summation of all momentum modes
in the ($1+1$) dimensional spacetime. $\eta_{\left(  \pm,k\right)  }=\int
\frac{\lambda d\tau}{\sqrt{(2\pi)^{3}2\omega}}e^{i\Omega\tau\pm ik^{\mu}%
x_{\mu}-\frac{\tau^{2}}{2\sigma_{\tau}^{2}}}$where $k^{\mu}x_{\mu}=\omega
t(\tau)-kx(\tau)$ are related to the motion of the atom. The ladder operators
$\sigma^{\pm}$ of the atom are defined by
\begin{align}
&  \sigma^{+}|e\rangle=0,\sigma^{+}|g\rangle=|e\rangle,\nonumber\\
&  \sigma^{-}|g\rangle=0,\sigma^{-}|e\rangle=|g\rangle.
\end{align}
And the creation and annihilation operators of the field are defined by
\begin{align}
&  a_{k}^{\dag}|n\rangle_{k}=\sqrt{n_{k}+1}|n+1\rangle_{k},\nonumber\\
&  a_{k}|n\rangle_{k}=\sqrt{n_{k}}|n-1\rangle_{k}.
\end{align}
where $|n\rangle_{k}$ is the Fock state of the field and indicates that there
are $n$ photons in the $k$ mode.

Hence the interaction between the accelerated atom and the field is described
according to%
\begin{align}
U|g\rangle|n\rangle_{k}=  &  |g\rangle|n\rangle_{k}-i\sqrt{n}\eta
_{(-,k)}|e\rangle|n-1\rangle_{k}\nonumber\\
&  -i\sqrt{n+1}\eta_{(+,k)}|e\rangle|n+1\rangle_{k},\label{e1}\\
U|e\rangle|n\rangle_{k}=  &  |e\rangle|n\rangle_{k}-i\sqrt{n+1}\eta
_{(-,k)}^{\ast}|g\rangle|n+1\rangle_{k}\nonumber\\
&  -i\sqrt{n}\eta_{(+,k)}^{\ast}|g\rangle|n-1\rangle_{k}. \label{e2}%
\end{align}
In the two equations (\ref{e1}) and (\ref{e2}), the second and third terms on
the right side of each equation represent the rotating-wave and
counter-rotating terms, respectively. For example, the second term on the
right side of Eq. (\ref{e1}) indicates that the atom absorbs one photon in the
field and jumps from the ground state to the excited state, which is also
called the stimulated absorption term. The third term on the right side of Eq.
(\ref{e1}) represents the contribution of the Unruh effect, which is generated
by the accelerated motion of the atom in the electromagnetic field and is
called the stimulated Unruh effect term. The Eq. (\ref{e2}) can be interpreted similarly.

At first, we take the field as the vacuum, i.e., there are no particles in the
field, or $n_{k}=0$ for any $k$ mode. When the atom is accelerated in the
vacuum with the trajectory \cite{takagi1986vacuum,ben2019unruh},
\begin{equation}
x^{\mu}(\tau)=[\sinh(a\tau)/a,0,0,\cosh(a\tau)/a], \label{ua}%
\end{equation}
we have the transition probability $P$ as%
\begin{equation}
P\propto\frac{B}{e^{2\pi\Omega/a}-1},
\end{equation}
where $B$ is a coefficient related to the initial conditions of the atom. This
represents a Bose-Einstein distribution at temperature $T=\frac{a}{2\pi}$, and
indicates the uniformly accelerated detector in the vacuum perceives an
apparently thermal field \cite{dewitt1975quantum,wald1994quantum}.

When the atom at the ground state is initially accelerated in the
electromagnetic field, the atom will jump from the ground to the excited state
in two different ways (i.e., the stimulated absorption and the stimulated
Unruh effect). From Eq. (\ref{e1}), it is not hard to calculate the
acceleration-induced transition probability as,
\begin{equation}
P_{a}\propto\frac{(n+1)B}{e^{2\pi\Omega/a}-1}, \label{pt}%
\end{equation}
where the field is taken as the single-mode Fock state $|n\rangle$. This
result indicates that when the atom is accelerated in the $n$-photon field,
its transition probability will be amplified by $n+1$ times compared with that
accelerated in the vacuum.

\section{Stimulated anti-Unruh effect}

In the previous section, it is shown that a detector accelerated in the
electromagnetic field is more sensitive to acceleration-induced thermal
effects than a detector accelerated in the vacuum field. It is mainly embodied
in the amplification of the transition probability of the atom. Now we
calculate the transition probability precisely and discuss its change with the acceleration.

Consider that an atom as the UDW detector is initially in the ground state
$|g\rangle$. When the atom is accelerated in the $k$-mode photon field,
according to Eq. (\ref{e1}), the probability of the transition to the excited
state is calculated as
\begin{equation}
P_{|g\rangle\rightarrow|e\rangle}=n_{k}|\eta_{(-,k)}|^{2}+(n_{k}%
+1)|\eta_{(+,k)}|^{2}, \label{PN}%
\end{equation}
where the first term on the right side is the contribution of the stimulated
absorption term and the second term is the contribution of the stimulated
Unruh effect term, as explained for the second and third terms on the right
side of Eqs. (\ref{e1}) and (\ref{e2}).

In order to present the influence of the field on the detector, we choose a
general initial state
\begin{equation}
\rho_{in}=|g\rangle\langle g|\otimes\sum_{n_{k},n_{k}^{\prime}}p_{n_{k}%
}p_{n_{k}^{\prime}}^{\ast}|n_{k}\rangle\langle n_{k}^{\prime}|,
\end{equation}
where $\sum_{n_{k},n_{k}^{\prime}}p_{n_{k}}p_{n_{k}^{\prime}}^{\ast}%
|n_{k}\rangle\langle n_{k}^{\prime}|$ represents the state of the
electromagnetic field, and $p_{n_{k}}$ represents the number distribution in
the $k$-mode. Thus, the transition probability becomes
\begin{align}
P=  &  \sum_{n_{k}}|p_{n_{k}}|^{2}n_{k}|\eta_{(-,k)}|^{2}+\sum_{n_{k}%
}|p_{n_{k}}|^{2}(n_{k}+1)|\eta_{(+,k)}|^{2}\nonumber\\
&  +\sum_{n_{k}}p_{n_{k}}^{\ast}p_{n_{k}+2}\sqrt{(n_{k}+1)(n_{k}+2)}%
\eta_{(-,k)}\eta_{(+,k)}^{\ast}\nonumber\\
&  +\sum_{n_{k}}p_{n_{k}}p_{n_{k}+2}^{\ast}\sqrt{(n_{k}+1)(n_{k}+2)}%
\eta_{(-,k)}^{\ast}\eta_{(+,k)}. \label{P}%
\end{align}
It is easily confirmed that the transition probability decays to the form in
Eq. (\ref{PN}) when the field state is taken as the Fock state. Eq. (\ref{P})
is a more general result, with which we can study the influence of different
field states. Note that some special states such as entangled states or
squeezed states \cite{luo2017deterministic,pan2020influence} can improve the
accuracy of experimental measurements. Therefore, it is interesting to
calculate the transition probability of the accelerated atom for the different
field states, and in the following we will take the photon states as the
coherent state, the thermal state, and the squeezed state, respectively.

When the field is taken as the single-mode coherent state
\cite{glauber1963coherent},
\begin{equation}
|\alpha\rangle=e^{-\frac{|\alpha|^{2}}{2}}\sum_{n_{k}=0}^{\infty}\frac
{\alpha^{n_{k}}}{\sqrt{n_{k}!}}|n_{k}\rangle,
\end{equation}
the transition probability in Eq. (\ref{P}) is calculated as
\begin{align}
P_{CS}  &  =|\alpha|^{2}|\eta_{(-,k)}|^{2}+(|\alpha|^{2}+1)|\eta_{(+,k)}%
|^{2}\nonumber\\
&  +\alpha^{2}\eta_{(-,k)}\eta_{(+,k)}^{\ast}+(\alpha^{\ast})^{2}\eta
_{(-,k)}^{\ast}\eta_{(+,k)}.
\end{align}
where $|\alpha|^{2}$ means the average number of photons.

When the field is taken as the thermal state, its expression is given as
\begin{equation}
\rho_{k}=\sum_{n_{k}}\left(  1-e^{-\beta\omega_{k}}\right)  e^{-n_{k}%
\beta\omega_{k}}|n_{k}\rangle\langle n_{k}|, \label{thermal}%
\end{equation}
where $\beta=1/T$ and $T$ is the temperature of the thermal field. The
evolution of the accelerated atom in such a field is similar to that studied
earlier in Ref. \cite{costa1995background,dai2015killing}, and the expected
value of the particle number operator can be calculated as
\begin{equation}
\langle n_{k}\rangle=tr(\rho_{k}n_{k})=\frac{1}{e^{\beta\omega_{k}}-1}.
\label{pno}%
\end{equation}
Substituting Eqs. (\ref{thermal}) and (\ref{pno}) into Eq. (\ref{P}), the
transition probability becomes
\begin{equation}
P_{TS}=\frac{1}{e^{\beta\omega_{k}}-1}|\eta_{(-,k)}|^{2}+(\frac{1}%
{e^{\beta\omega_{k}}-1}+1)|\eta_{(+,k)}|^{2}. \label{tsp}%
\end{equation}

When the field is taken as a single-mode squeezed state
\cite{ma2011quantum,lo1993generalized}, we have the expression
\begin{equation}
|S_{k}\rangle=\frac{1}{\sqrt{\cosh r_{k}}}\sum_{n_{k}=0}^{\infty}\left(
-e^{i\phi}\tanh r_{k}\right)  ^{n_{k}}\frac{\sqrt{(2n_{k})!}}{2^{n_{k}}n_{k}%
!}|2n_{k}\rangle.
\end{equation}
Such a state can be generated by implementing the squeezing operator
$S(\zeta)=\exp\left[  \frac{1}{2}\left(  \zeta^{\ast}\hat{a}_{k}^{2}-\zeta
\hat{a}_{k}^{\dagger2}\right)  \right]  $ on the vacuum state, $|S_{k}%
\rangle=S(\zeta)|0\rangle$, where $\zeta=re^{i\phi}$, $r$ is squeezing
parameter and $\phi$ is the spatially azimuthal angle. For simplicity, we take
$\phi=0$ in the following calculation. The average number of photons in this
squeezed state is obtained as
\begin{equation}
\langle n_{k}\rangle_{S_{k}}=\sinh^{2}(r_{k}). \label{an}%
\end{equation}
So the transition probability for the accelerated atom in such a field state
is calculated as
\begin{align}
P_{SS}=  &  \sinh^{2}(r_{k})|\eta_{(-,k)}|^{2}+[\sinh^{2}(r_{k})+1]|\eta
_{(+,k)}|^{2}]\nonumber\\
&  +\sum_{n_{k}}Q_{n_{k}}[\eta_{(-,k)}\eta_{(+,k)}^{\ast}+\eta_{(-,k)}^{\ast
}\eta_{(+,k)}]. \label{ssp}%
\end{align}
where $Q_{n_{k}}=p_{n_{k}}^{\ast}p_{n_{k}+2}\sqrt{(n_{k}+1)(n_{k}+2)}$, and
$p_{n_{k}}=\frac{1}{\sqrt{\cosh(r_{k}))}}[\tanh(r_{k})]^{n}\frac{\sqrt
{(2n_{k})!}}{2^{n_{k}}n_{k}!}$is the distribution of the single-mode squeezed
state under the base of Fock states.

\begin{figure}[ptb]
\includegraphics[width=1\linewidth]{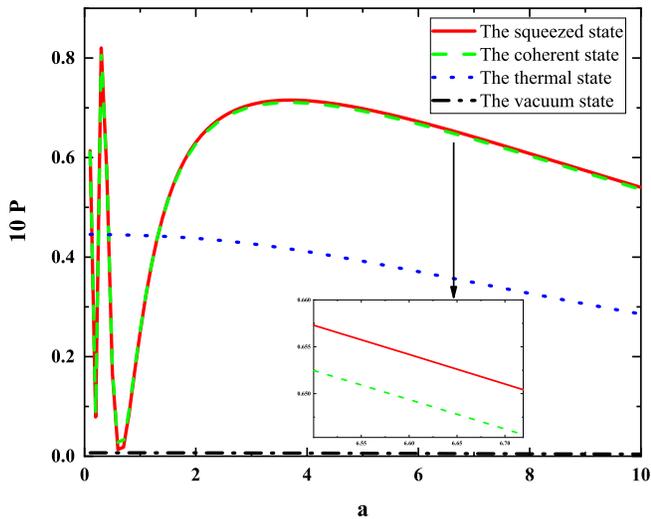} \caption{The transition probability
as a function of the acceleration $a$. The red solid line denotes the case
that the field state is taken as a squeezed state. The green dashed line
denotes the case that the field state is taken as the coherent state. The blue
dotted line denotes the case that the field state is taken as the thermal
state. The black dotted dashed line denotes the case that the field state is
taken as the vacuum state. The model parameters employed are $\varsigma=0.3$,
$m=k=1$, $\Omega=1$, $\langle n\rangle=29.1292$.}%
\label{Fig1}%
\end{figure}

Fig. 1 presents the change of different transition probabilities corresponding
to different field states with the acceleration of the atom. It is found that
the transition probability corresponding to every field state decreases when
the acceleration is increased, which is an evident phenomenon that confirms
the existence of the anti-Unruh effect
\cite{brenna2016anti,garay2016thermalization}. The reason for the fluctuation
at small acceleration in Fig. 1 is that the contribution from the light-matter
interaction dominates and the acceleration-induced effect is small there. On
the other hand, it is also evidently seen that the transition probability
corresponding to the coherent, thermal or squeezed state is indeed enhanced
compared with the case when the atom is accelerated in the vacuum. In
particular, the enhanced effect for the squeezed state is better than that for
any other state under the condition that the average atomic number is the same
for every field state.

\begin{figure}[ptb]
\centering
\includegraphics[width=1\linewidth]{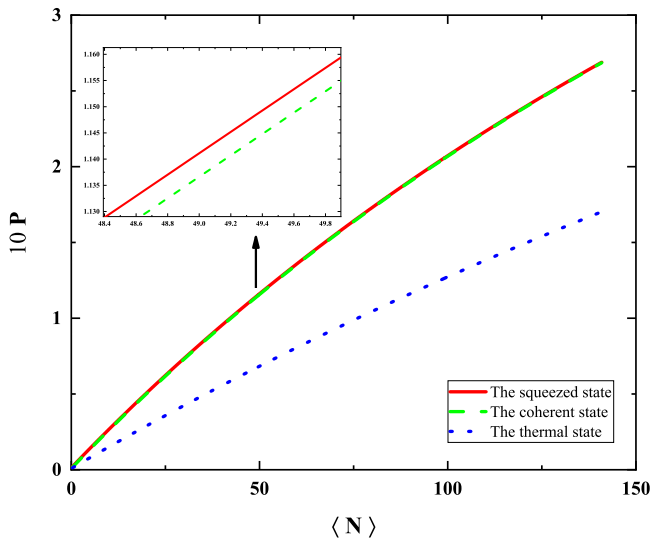}\caption{The transition probability
as a function of the number of particles. The acceleration takes $a=4$, and
the other parameters take the same values as in Fig.1 }%
\label{Fig2}%
\end{figure}

For the purpose of observation, we present the change of the transition
probability with the average number of particles, as given in Fig. 2. It is
seen that the amplification of the transition probability is more obvious for
the squeezed state than that for any other state as the field state. The
stimulated acceleration-induced effect can be detected via an accelerated
low-mass two-level system \cite{svs2021}. When the field state is taken as the
vacuum state, the spontaneous transition probability is only $P_{0}%
\approx10^{-18}$ which is obtained by considering the accelerated electrons in
storage rings with the acceleration $a\approx10^{17}ms^{-2}$ \cite{bl1987}.
According to our calculation, the transition probability can increase about
$n$ times when the field is taken as the electromagnetic field, i.e., the
transition probability becomes $2.34305\times10^{-14}$ for the squeezed state
as the field state, where $n=1000$, other parameters take the same values as
that in the acceleration of electrons \cite{bl1987}, and all the physical
constants such as the light velocity $c$, the reduced Planck constant $\hbar$,
and the Boltzmann constant $k_{B}$ are put into the Eq. (\ref{ssp}). While in
the actual experiment, the phonon number can reach $10^{15}$
\cite{hsp2014,rjs2016}, which implies that the possibility of observing the
acceleration-induced effect under the present experimental conditions.

Moreover, from Eq. (\ref{P}), it is noted that the acceleration affects not
only the stimulated Unruh effect term but also the stimulated absorption term.
So the phenomenon presented in Fig. 1 is not completely from the contribution
of the acceleration, and the usual absorption of the atom from the field works
in the process. We check the contribution of the stimulated absorption term
for the transition of the atom from the ground to excited state with the
proportional parameter, $\Lambda=\frac{\sum_{n_{k}}p_{n_{k}}n_{k}|\eta
_{(-,k)}|^{2}}{P}$. It is found that the value of $\Lambda$ for every field
state decreases as the acceleration increases, and the contribution of the
stimulated absorption terms to the transition probability is larger for the
squeezed state than that for any other state. Although the influence of the
stimulated absorption term is reduced by increasing the acceleration of the
atom, the transition probability also decreases in the process. Thus, it
cannot deduce directly that the contribution from the stimulated Unruh effect
term decreases as the acceleration increases. So it requires a specific method
in which the absorption term can be suppressed to confirm the existence of the
anti-Unruh effect.

\section{Acceleration-induced transparency}

The recent study in Ref. \cite{vsoda2022acceleration} provides a method called
acceleration-induced transparency, which can suppress the contribution of the
stimulated absorption term for the transition of an accelerated atom in the
electromagnetic field. In this section, we will discuss whether it is also
feasible for the appearance of the anti-Unruh effect.

The acceleration-induced transparency means that in the process of interaction
between the accelerated atom and the field, the rotating-wave term can be much
suppressed, and the counter-rotating term dominates in the transition of the
atom
\cite{obadia2007unruh,kothawala2010response,reagor2016quantum,ahmadzadegan2018unruh}%
. This can be mathematically expressed as the condition that $\frac
{|\eta_{(-,k)}|}{|\eta_{(+,k)}|}\ll1$, where $\eta_{(-,k)}$ is related to the
rotating-wave term and $\eta_{(+,k)}$ is related to the counter-rotating term
as given in Eq. (\ref{e1}) and (\ref{e2}). For an accelerated atom at the
ground state initially, this condition means that the transition process of
the atom is caused mainly by the stimulated Unruh effect term and the
contribution from the stimulated absorption is much suppressed. In the
following, we will discuss how to realize this condition. In order to present
it clearly, we take the switching function as $\chi(\tau)=1$.

Now we introduce a phase function $\alpha(\tau)=k^{\mu}x_{\mu}(\tau)$
determined by the atomic world line $x_{\mu}(\tau)$, and the transition
amplitude can be rewritten as
\begin{equation}
\eta_{\pm,k}=\int qd\tau e^{i\Omega\tau\pm i\int_{0}^{\tau}\dot{\alpha}%
(\tau^{\prime})d\tau^{\prime}},
\end{equation}
where $q$ is a coefficient determined by the initial detector settings and the
field frequency \cite{crispino2008unruh}. For the conventional Unruh effect,
the world line of a uniformly accelerated detector is given by Eq. (\ref{ua})
and $k^{\mu}=\{1,0,0,1\}$ is taken for the excited particles associated with
the uniformly accelerated detector in the vacuum without loss of generality
\cite{ahmadzadegan2018unruh,giacomini2022second}. Thus, we have,
$\eta_{(+,\Omega)}=\frac{-iq}{4a\pi\sqrt{\pi}}e^{\frac{-i}{a}}e^{i\frac
{\Omega}{a}\ln\left(  \frac{-i}{a}\right)  }\Gamma(-i\Omega/a)$ where $\Gamma$
is the complex Gamma function. For our purpose, a non-uniformly accelerated
world line is required as,
\begin{equation}
\dot{\alpha}(\tau):=k\left\{
\begin{array}
[c]{ll}%
v_{0} & \tau<0\\
v_{0}+a_{1}\tau & \tau\in\left[  0,T_{1}\right)  \\
v_{1}+a_{2}\left(  \tau-T_{1}\right)   & \tau\in\left[  T_{1},T_{2}\right)  \\
v_{2} & \tau\geq T_{2}%
\end{array}
,\right.  \label{na}%
\end{equation}
as in Ref. \cite{vsoda2022acceleration}.

\begin{figure}[ptb]
\centering
\includegraphics[width=1\linewidth]{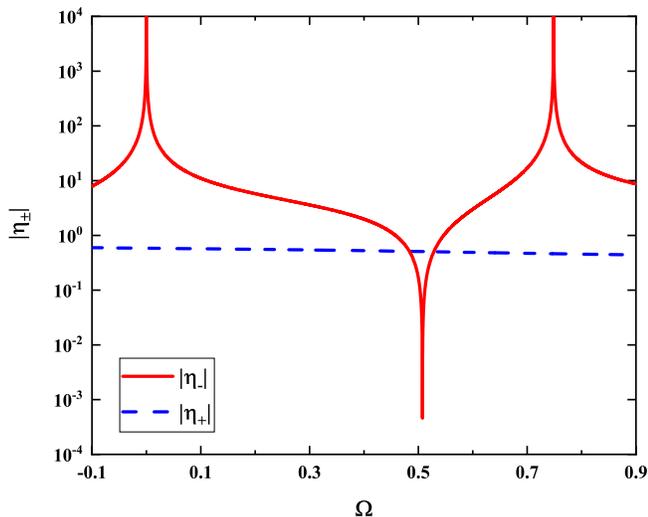} \caption{The transition amplitude as
a function of the atomic energy gap. The corresponding parameters are taken as
$v_{0}=1.041$, $v_{1}=1.07$, $v_{2}=1.05141$, $T_{1}=9.7435$, $T_{2}%
=1305.412$, $\omega=k=1$. The trajectory of the atom is given in Eq.
(\ref{na}). }%
\label{Fig3}%
\end{figure}

From this expression in Eq. (\ref{na}), it is seen that the atom moves
inertially at velocity $\omega-v_{0}$ at the initial time. The first
acceleration process is implemented with the acceleration $a_{1}=(v_{1}%
-v_{0})/T_{1}$, where the velocity becomes $\omega-v_{1}$ at the end of this
process. Then, the second acceleration process is started with the
acceleration $a_{2}=(v_{2}-v_{1})/(T_{2}-T_{1})$. Finally, the motion of the
atom remains at the velocity of $\omega-v_{2}$. In Fig. 3, we plot $\eta
_{\pm,k}$ for such a trajectory and obtain a similar result to that in Ref.
\cite{vsoda2022acceleration}. It shows clearly that at some specific
frequencies, the rotating-wave term represented by $\eta_{(-,k)}$ is much suppressed.

However, it is not easy to determine whether the anti-Unruh effect appears for
the above process, since one requires checking the change of the transition
probability with the acceleration (but there are two accelerations in the
process described above) for the determination of the anti-Unruh effect. Thus,
we have to choose another accelerated world line for the atom, given as,
\begin{equation}
\dot{\alpha}(\tau):=k\left\{
\begin{array}
[c]{ll}%
v_{0} & \tau<0\\
v_{0}+a\tau & \tau\in\left[  0,T\right) \\
v & \tau\geq T,
\end{array}
\right.  \label{aa}%
\end{equation}
where $a=(v-v_{0})/T$ represents the acceleration in the accelerated process.

\begin{figure}[ptb]
\centering
\includegraphics[width=1\linewidth]{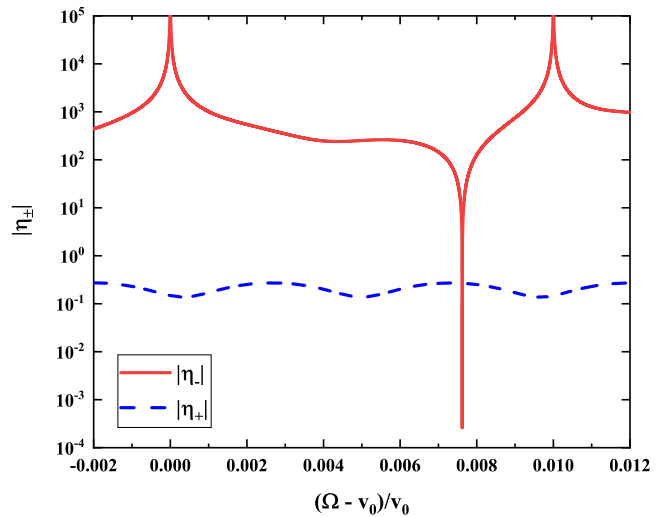} \caption{The transition amplitude as
a function of atomic energy gap. The corresponding parameters are taken as
$v_{0}=1.03$, $T=5.105$, $v=1.801$, $\omega=k=1$. The trajectory of the atom
is given in Eq. (\ref{aa}).}%
\label{Fig4}%
\end{figure}

With this accelerated world line (\ref{aa}), we study whether the stimulated
absorption term can be suppressed. Fig. 4 gives the corresponding results. It
shows that the transition amplitude caused by the stimulated absorption term
is highly suppressed by choosing a specific energy gap, e.g., it is
$\Omega=0.51$ as presented in Fig. 4. Thus, the phenomenon of
acceleration-induced transparency occurs for the uniformly accelerated motion.
Since the stimulated absorption term is suppressed and the stimulated Unruh
effect term dominates for a specific energy gap of the atom, we can discuss
whether the anti-Unruh effect exists with such an atomic energy gap. The
expression of the transition probability in Eq. (\ref{P}) is rewritten as

\begin{figure}[ptb]
\centering
\includegraphics[width=\linewidth]{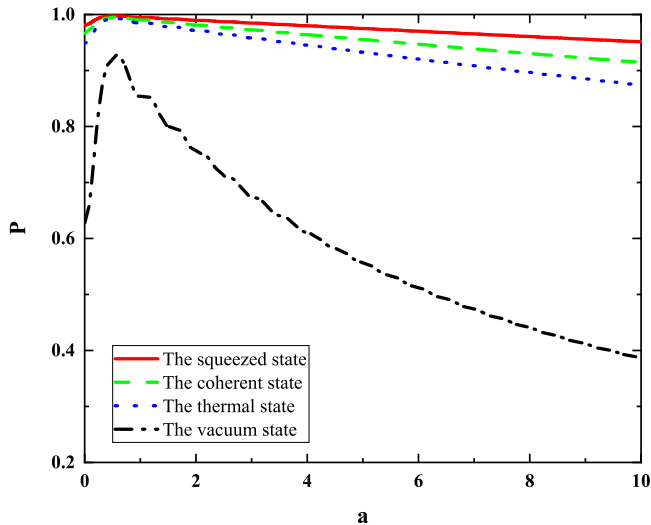} \caption{The transition probability
as a function of acceleration $a$ for an accelerated atom with acceleration
part of the world line (\ref{aa}). The corresponding parameters are taken as
$v_{0}=1.03$, $T=5.105$, $\Omega=0.51$, $k=1$, $\beta=0.06728$, $\alpha=4$,
$r=2.4$.}%
\label{Fig5}%
\end{figure}%

\begin{equation}
P=\sum_{n_{k}}|p_{n_{k}}|^{2}(n_{k}+1)|\eta_{(+,k)}|^{2}, \label{P one}%
\end{equation}
where $\eta_{(-,k)}$-related terms have infinitesimal values and are neglected
here. Fig. 5 presents the anti-Unruh effect for the different field states
when the energy difference takes $\Omega=0.51$ where the stimulated absorption
term is suppressed. Note that the acceleration part of the detector's world
line (\ref{aa}) is only considered here since the inertial parts will disturb
the judgement on the existence of the anti-Unruh effect. In particular, the
slight rising of the curve at the small acceleration in Fig. 5 is due to the
integral of the finite time for calculating $\eta_{(+,k)}$. As we have
checked, if the time for the integral in $\eta_{(+,k)}$ in expanded to
infinity, the rising phenomena at the small acceleration in Fig. 5 will
disappear. It is seen in Fig. 4 that the anti-Unruh effect can be enhanced by
the stimulated interaction process under the condition that
acceleration-induced transparency happens. According to our calculation, more
enhancement is realized using the squeezed field states than any other field
state. From Eq. (\ref{P one}), it is not hard to see that the transition
probability is increased approximately $n$ times when the atom is accelerated
in the electromagnetic field, compared with the case in which the atom is
accelerated in the vacuum. As the analyses made for Fig. 2, the observation is
possible under the present experimental conditions, but it deserves further
study about how to implement the experiment. For our discussion here, the
anti-Unruh effect and the acceleration-induced transparency can appear at the
same time, although each one of them requires a specific energy level gap to
be realized.

\section{Conclusion}

In this paper, we investigate the transition process of a two-level atom as
the UDW detector accelerated in the electromagnetic field. We calculate the
transition probability of the accelerated atom initially at the ground state
for different field states, including the Fock state, the coherent state, the
thermal state, and the squeezed state. The transition probability is enhanced
even when the anti-Unruh effect happens. Meanwhile, it is found that the
enhancement phenomenon is more prominent by choosing the squeezed state than
by choosing any other state as the field state. However, this enhancement
includes the contribution from the stimulated Unruh effect term and the
stimulated absorption term. The latter one can disturb the estimation for the
existence of anti-Unruh effect. We select a specific trajectory for the
accelerated atom to suppress the contribution to the transition probability
from the stimulation absorption term. This is just the recently found
acceleration-induced transparency if the stimulation absorption term can be
much suppressed for the atoms with a specific energy difference. When the
acceleration-induced transparency is realized, we find that the anti-Unruh
effect still appears, which shows that the stimulated interaction between the
accelerated atom and the field can enhance the possibility of observing the
anti-Unruh effect experimentally. Since the anti-Unruh effect leads to some
different phenomena (e.g., entanglement is increased with the increasing
acceleration) from that caused by the Unruh effect or the usual thermal effect
\cite{li2018would,pan2020influence,pan2021thermal}, it is more advantageous to
detect the anti-Unruh effect in the future feasible experiment using the
squeezed state as the background field.

\section{Acknowledge}

This work is supported by the NSFC grant Nos. 11654001.

\end{document}